\documentclass[aps,prl,twocolumn,groupedaddress]{revtex4}

\usepackage{epsfig}

\begin{document}

\def\d{{\rm d}}
\def\p{I\!\!P}

\def\lp{\left. }
\def\rp{\right. }
\def\lr{\left( }
\def\rr{\right) }
\def\le{\left[ }
\def\re{\right] }
\def\lg{\left\{ }
\def\rg{\right\} }
\def\lb{\left| }
\def\rb{\right| }

\def\beq{\begin{equation}}
\def\eeq{\end{equation}}
\def\bea{\begin{eqnarray}}
\def\eea{\end{eqnarray}}

\preprint{LPSC 04-069}
\title{Evidence for Factorization Breaking in Diffractive Low-$Q^2$ Dijet Production}
\author{Michael Klasen}
\email[]{klasen@lpsc.in2p3.fr}
\affiliation{Laboratoire de Physique Subatomique et de Cosmologie,
 Universit\'e Joseph Fourier/CNRS-IN2P3, 53 Avenue des Martyrs,
 F-38026 Grenoble, France}
\author{Gustav Kramer}
\affiliation{{II.} Institut f\"ur Theoretische Physik, Universit\"at
 Hamburg, Luruper Chaussee 149, D-22761 Hamburg, Germany}
\date{\today}
\begin{abstract}
We calculate diffractive dijet production in deep-inelastic scattering
at next-to-leading order of perturbative QCD, including contributions from
direct and resolved photons, and compare our predictions to preliminary data
from the H1 collaboration at HERA. In contrast to recent experimental
claims, evidence for factorization breaking is found only for resolved, and
not direct, photon contributions. No evidence is found for large
normalization uncertainties in diffractive parton densities. The results
confirm theoretical expectations for the (non-)cancellation of soft
singularities in diffractive scattering as well as previous results for
(almost) real photoproduction.
\end{abstract}
\maketitle

\begin{figure}
 \centering
 \includegraphics[width=\columnwidth]{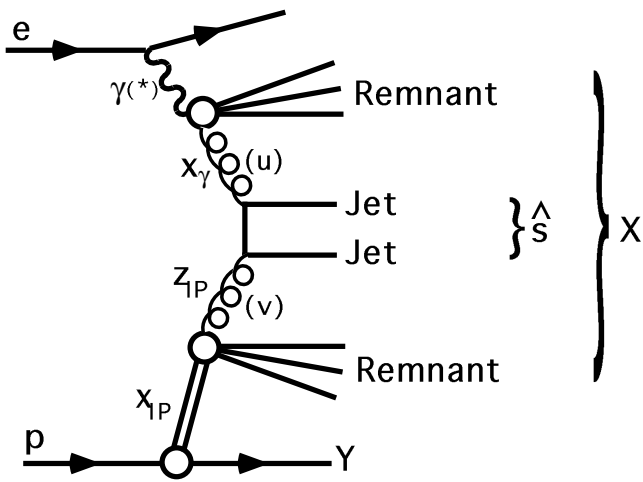}
 \caption{\label{fig:1}Diffractive scattering process $ep\to eXY$, where
 the hadronic systems $X$ and $Y$ are separated by the largest rapidity
 gap in the final state.}
\end{figure}

At current high-energy colliders such as HERA or the TEVATRON,
diffractive processes are known to constitute an important fraction of all
scattering events. These events are defined experimentally by the presence
of a forward-going hadronic system $Y$ with four-momentum $p_Y$, low mass
$M_Y$ (typically a proton that has remained intact), small four-momentum
transfer $t=(p-p_Y)^2$, and small longitudinal momentum transfer $x_{\p}=
q.(p-p_Y)/(q.p)$ from the incoming proton with four-momentum $p$ to the
central hadronic system $X$ (see Fig.\ \ref{fig:1}).
%
%
The presence of a hard scale, such as the squared photon virtuality $Q^2=
-q^2$ in deep-inelastic scattering (DIS) or large transverse jet momenta
$p_{T,~{\rm jet~1,2}}^*$ in the photon-proton center-of-momentum frame,
should then allow for calculations of partonic cross sections for the
central system $X$ using perturbative QCD. \\

The question whether diffractive cross sections are factorizable, like
inclusive cross sections, into universal, diffractive parton densities
$f_i^D$ and partonic cross sections $\sigma_{\gamma^*i}$,
\bea
 {\d^2\sigma\over\d x_{\p}\ \d t}\!&\!=\!&\!\int_x^{x_{\p}}\!\!\d\xi\
 f_i^D(\xi,Q^2,x_{\p},t)\ \sigma_{\gamma^*i}(x,Q^2,\xi),
\eea
is at the heart of the current debate in diffractive physics and may even
affect the prospects for discovery of new particles such as the Higgs boson
at the LHC \cite{Khoze:2000cy,Boonekamp:2001vk}.\\

For DIS processes, factorization has indeed been proven to hold
\cite{Collins:1997sr}, and diffractive parton densities have been extracted
from HERA data at low and intermediate $Q^2$ \cite{h1ichep02,
Chekanov:2004hy,Martin:2004xw}. However,
factorization breaks down for processes with two initial-state hadrons,
where not all singularities from soft gluon exchange cancel
\cite{Collins:1997sr} and initial state rescattering effects become
important \cite{Kaidalov:2001iz}. At the Tevatron, this leads to a
suppression of the expected dijet rate by about a factor of ten
\cite{Affolder:2000vb}. \\

Processes with (almost) real photons are unique in the respect that they
involve direct interactions of the photon with partons from the proton as
well as resolved photon contributions leading to parton-parton interactions
(for a recent review see \cite{Klasen:2002xb}). As in the case of DIS,
factorization should hold for direct photoproduction, whereas it is expected
to fail for the hadron-like
resolved processes. A two-channel eikonal model using vector-meson dominated
photon fluctuations predicts a suppression by about a factor of three for
resolved photoproduction at HERA \cite{Kaidalov:2003xf}. This suppression
factor has recently been applied to diffractive dijet production
\cite{Klasen:2004tz,Klasen:2004qr} and compared to preliminary data from H1
\cite{h1ichep04} and ZEUS \cite{zeusichep04}. While at leading order (LO) of
perturbative QCD no suppression of the resolved contribution seems
necessary, next-to-leading order (NLO) corrections, calculated previously
for inclusive direct \cite{Klasen:1995ab} and resolved \cite{Klasen:1996it}
dijet photoproduction, increase the cross sections significantly
\cite{Klasen:2004tz,Klasen:2004qr} and require indeed a suppression factor,
showing that factorization breaking occurs at this order for {\em resolved}
photoproduction.\\

Although a suppression factor is clearly not required theoretically for
{\em direct} photoproduction, the H1 \cite{h1ichep04} and ZEUS
\cite{zeusichep04} collaborations have recently shown their data to be
consistent with a {\em global} suppression of about a factor of two for
direct {\em and} resolved NLO QCD predictions, claiming evidence for
factorization breaking also in {\em direct} photoproduction \cite{h1ichep04}
and citing uncertainties from diffractive parton densities
\cite{zeusichep04}. These claims were based on the observation that the
measured distributions in the observed photon momentum fraction
$x_\gamma^{\rm obs}$ and in the kinematically related variable $y$ were
better described by a global than by a resolved-only suppression factor.
However, these variables are known to be subject to large hadronization
corrections at small $p_T^*$ \cite{hadcorr}. In fact a global
suppression factor underestimates the data at large $p_T^*$, where direct
photoproduction dominates \cite{h1ichep04}.\\

Obviously, these contradicting theoretical and experimental interpretations
require clarification from an independent physical analysis. A particularly
suitable physical process is low-$Q^2$ electroproduction of dijets,
{\em i.e.} the transition region of (almost) real photoproduction to DIS.
Resolved processes are known to continue to be important at low, but finite
$Q^2$, where the perturbative quark-antiquark splitting induces large higher
order QCD corrections that have to be resummed due to the presence of a
large mass logarithm. Consequently, any possible suppression for resolved
photons should only show up at low, but not high $Q^2$, whereas a global
suppression should be visible for {\em all} $Q^2$.\\

A consistent factorization scheme for virtual photoproduction has been
defined and the full (direct and resolved) NLO corrections for inclusive
dijet production have been calculated in \cite{Klasen:1997jm}.
In this Letter, we therefore adapt our {\em inclusive} NLO calculation to
{\em diffractive} dijet production at low $Q^2$ and compare our predictions
with the corresponding data from H1 \cite{Schatzel:2004be} in order to study
the turn-on of factorization breaking as the photon virtuality decreases
from the DIS to the photoproduction region. For this purpose, integrations
over the additional diffractive variables $t$ and $x_{\p}$ and the
kinematics corresponding to the H1 experiment have been implemented in the
NLO Monte Carlo program JetViP \cite{Potter:1999gg}. In particular, we
use an electron and proton beam energy of 27.5 and 820 GeV, respectively, as
well as the following kinematical ranges: $4$ GeV$^2<Q^2<80$ GeV$^2$,
electron momentum transfer $0.1<y<0.7$, $x_{\p}<0.05$, $|t|<1$ GeV$^2$,
$M_Y<1.6$ GeV, $p_{T,~{\rm jet~1,2}}^*>5 (4)$ GeV, and jet rapidities
$-3<\eta^*_{\rm jet~1,2}<0$. Jets are defined by the CDF cone algorithm
with jet radius $R=1$.\\

While the original H1 analysis used a symmetric cut of 4 GeV on the
transverse momentum of both jets \cite{Adloff:2000qi}, asymmetric cuts are
required for infrared-stable comparisons to NLO calculations
\cite{Klasen:1995xe}. The data have therefore been reanalyzed for asymmetric
cuts \cite{Schatzel:2004be}, using the LO Monte Carlo program RAPGAP
\cite{Jung:1993gf}, and compared to a NLO calculation for diffractive dijet
production in DIS \cite{Hautmann:2002ff}. In the extraction of the
diffractive parton densities, the usual Regge factorization formula
\cite{Ingelman:1984ns}
\bea
 f_i^D(\xi,Q^2,x_{\p},t)&=&f_{\p/p}(x_{\p},t) f_{i/\p}(\xi/x_{\p},Q^2).
\eea
has been employed. For our LO and NLO resolved virtual photon predictions,
we have used the parton densities SaS1D \cite{Schuler:1996fc} at the fixed
factorization scale 40 GeV and transformed them from the DIS$_\gamma$ to the
$\overline{\rm MS}$ scheme \cite{Klasen:1997jm}.\\

\begin{figure}
 \centering
 \includegraphics[width=\columnwidth]{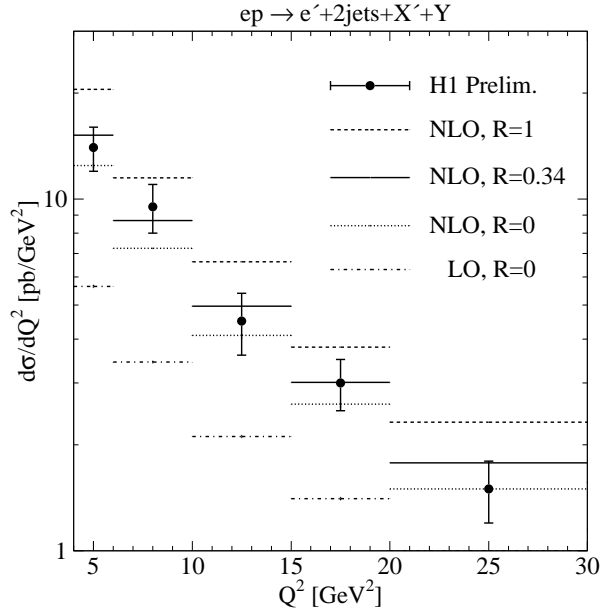}
 \caption{\label{fig:2}Dependence of the diffractive dijet cross section at
 HERA on the squared photon virtuality $Q^2$. Preliminary H1 data
 \cite{Schatzel:2004be} are compared with theoretical predictions for DIS at
 LO (dot-dashed) and NLO (dotted) and NLO predictions including resolved
 virtual photon contributions with (full) and without (dashed curve) a
 suppression factor of $R=0.34$.}
\end{figure}

We now turn to our numerical results. The dependence of the diffractive
dijet cross section at HERA on the squared photon virtuality $Q^2$ is shown
in Fig.\ \ref{fig:2}, where the preliminary H1 data \cite{Schatzel:2004be}
%
%
%
\begin{figure*}
 \centering
 \includegraphics[width=0.836\textwidth]{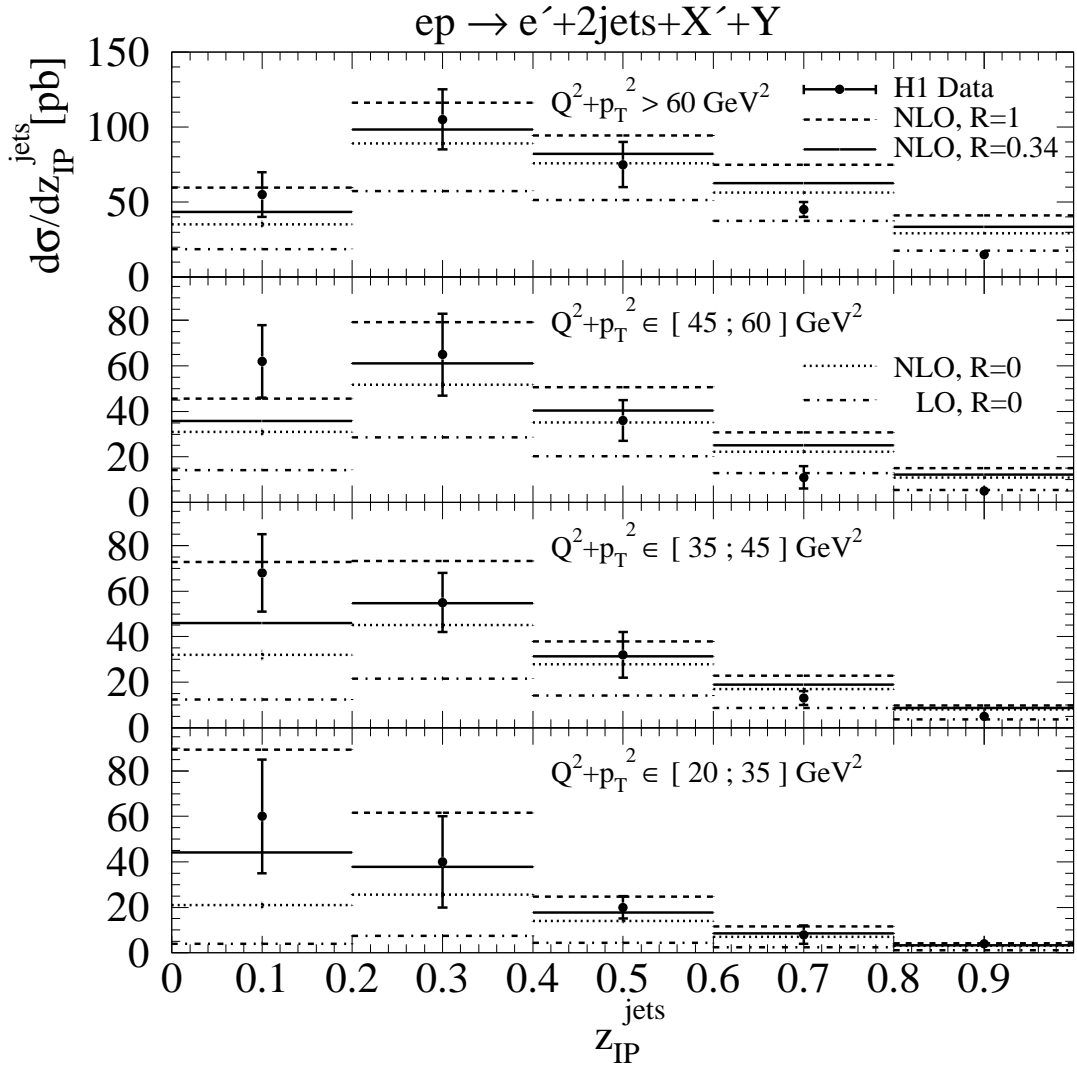}
 \caption{\label{fig:3}Dependence of the diffractive dijet cross section at
 HERA on the parton momentum fraction in the pomeron $z_{\p}$ for different
 ranges of $Q^2+p_T^{*2}$. Preliminary H1 data \cite{Schatzel:2004be} are
 compared with various theoretical predictions (see Fig.\ \ref{fig:2}).}
\end{figure*}
%
are compared with our theoretical predictions for DIS at LO (dot-dashed) and
NLO (dotted) as well as NLO predictions including resolved virtual photon
contributions with (full) and without (dashed curve) a suppression factor of
$R=0.34$. The LO and NLO DIS results agree with those obtained in
\cite{Hautmann:2002ff}, and the latter describe the data well in the
high-$Q^2$ region, where direct contributions should be sufficient. As
expected, the resolved virtual photon contributions increase the theoretical
predictions. They improve the description of the data at low $Q^2$, but only
if they are suppressed by the same factor of $R=0.34$ observed previously
for (almost) real photoproduction \cite{Klasen:2004tz,Klasen:2004qr}. Note
that the absolute normalization is now fixed for all $Q^2$ in a single
experiment, showing that neither factorization breaking in the direct
contributions nor large uncertainties from diffractive parton densities are
present. In particular, a common suppression factor of two for the direct
{\em and} resolved contributions would underestimate the data at all $Q^2$.
Furthermore, the presence of the large scale $Q^2$ in addition to
the relatively small scale $p_T^*$ leads to negligible hadronization
corrections \cite{Schatzel:2004be}.

Further evidence for the turn-on of factorization breaking in the transition
of high- to low-$Q^2$ diffractive dijet production can be derived from Fig.\
\ref{fig:3}, where the dependence of the cross section on the longitudinal
momentum fraction of the partons in the pomeron $z_{\p}$ is shown for
different ranges of $Q^2+p_T^{*2}$. Concentrating on the low-$z_{\p}$
regions,
we find that the NLO DIS predictions underestimate the data and need to be
supplemented at low $Q^2+p_T^{*2}$ by resolved virtual photon contributions,
which must again be suppressed by a factor of $R=0.34$ in order for the
predictions not to overestimate the data. As expected, the resolved
contributions become less important at high $Q^2+p_T^{*2}$, where the direct
and total NLO calculations become very similar and describe the data almost
equally well. The LO DIS prediction underestimates the data everywhere
except at high $Q^2$ and $z_{\p}$, where consequently all NLO predictions
overestimate the data. In this region, the agreement at NLO improves
considerably, if hadronization corrections are taken into account
\cite{Schatzel:2004be}.

In summary, we have presented a NLO analysis of diffractive dijet
production in low-$Q^2$ DIS. Contributions from direct and resolved photons
have been superimposed and compared with recent preliminary data from the
H1 collaboration at HERA. In contrast to experimental claims, evidence for
factorization breaking is found only for resolved, and not direct, photon
contributions, and no evidence is found for large normalization
uncertainties in the diffractive parton densities. The results confirm
theoretical expectations for factorization and its breaking in photon-hadron
and hadron-hadron scattering as well as previous results for almost real
photoproduction.

%
%



\end{document}